\documentstyle[11pt]{article}
\begin{document}

\def\thefootnote{\fnsymbol{footnote}}
\def\integer{\rm integer}
 \def\R#1{R^{#1}}
\def\pr#1{\phi_{#1}}
\def\mod{\rm mod}
 \def\s#1#2{{\tilde S_{#1,#2}}}
\def\draft{{\hskip 4cm {\Large DRAFT}}}

\def\B{ {\cal B}}
\def\L{ {\cal L}}
\def\V{ {\cal V}}
\def\S{ {\bf S}}
\def\H{ {\bf H}}
\def\R{ {\bf R}}
\def\Ref{ {\cal R}}
\def\W{\cal W}

\begin{center}

\vskip 0.5 cm

{\large \bf  Hyperbolic Billiards on Surfaces of Constant Curvature}

\vskip 1 cm

 Boris Gutkin  and Uzy Smilansky

\vskip 1 cm

{\em Department of Physics of Complex Systems\\
The Weizmann Institute of Science\\
Rehovot 76100\\
ISRAEL}\\{\small E-mail: fegutkin@vegas.weizmann.ac.il}
\vskip .5 cm
and
\vskip .5 cm
Eugene  Gutkin
\vskip 1 cm
{\em Department of Mathematics \\
University of Southern California\\
Los Angeles, CA 90089-1113\\
 USA}
\\{\small E-mail: egutkin@math.usc.edu}
\vskip 1 cm

\end{center}

\begin{abstract}

We establish sufficient conditions for the hyperbolicity of the billiard dynamics on surfaces of constant curvature. This extends known results for planar
billiards.
Using these conditions, we construct large classes of billiard
tables with positive Lyapunov exponents on the sphere and on the hyperbolic
plane.

\end{abstract}

\newpage



\section{ Introduction}                     

\noindent From the point of view of  differential dynamics,
 billiards are the geodesic flows on manifolds with a boundary.
Since the early beginnings of the study of  classical
 and quantum chaos, billiards have been used as a paradigm.
 Billiards are one of the best understood classes of dynamical systems that
demonstrate a broad variety of behaviors: from integrable to chaotic.
In fact, several  key properties of chaotic
dynamics were first observed and demonstrated for billiards.
 Many popular models of statistical mechanics, e.g., the Lorenz gas,
the hard sphere (Boltzmann - Sinai) gas, etc., can be reduced to 
billiards in
 special domains.

Among chaotic dynamical systems, the billiards with
 nonvanishing Lyapunov exponents are of special interest.
For brevity we will often call them hyperbolic billiards.  The Pesin theory
of smooth nonuniformly hyperbolic
systems [Pe], extended by A. Katok and J.-M. Strelcyn to systems with
singularities
[KS], implies that  hyperbolic billiards
have strong mixing properties: at most countable number of ergodic components,
positive entropy, Bernoulli property, etc.

In the present paper we consider billiards on surfaces of constant curvature.
For simplicity of exposition, we restrict the details of our analysis to the
simply connected surfaces of constant curvature: the plane, the sphere and
the hyperbolic plane. Employing a uniform method, we establish
widely applicable conditions, sufficient for positivity of the
Lyapunov exponent. The study of  billiards on curved surfaces is partially
motivated by recent technical  advances in semiconductor fabrication
techniques. They allow to manufacture solid
 state (mesoscopic) devices where electrons are confined to a
 curved surface (e.g. sphere) [FLBP]. Many properties of these devices can be
theoretically derived, using billiards as simplified models.

The billiard dynamics  crucially depends on the curvature
 of the surface. On the plane, billiard trajectories separate only linearly
 with time, so that the motion  between collisions with the
boundary is neutral. Exponential separation of billiard trajectories can
 occur only if the  reflections from the boundary introduce sufficient
 instability. On the hyperbolic plane, geodesics diverge exponentially,
so that the main role of the boundary is to confine the mass
 point to the billiard table. Thus, the boundary can be neutral (i. e.,
with zero
curvature), and the ``stretching and folding''
 necessary for chaotic dynamics,  will be provided
by the metric.  This phenomenon contrasts  the billiard dynamics on the
sphere,
where any two geodesics intersect twice, at focal points. Thus,
the boundary reflections have to compensate for the focusing effect
of the sphere, in order to produce chaotic dynamics.

Up to now, the study of billiards on surfaces (and hyperbolic billiard dynamics
in particular) has been by and large restricted to the Euclidean plane.
See, however, [Ve] for a study of  integrable billiards on surfaces of constant
nonzero curvature. See also [Ta] for some results on chaotic billiards on
 the hyperbolic plane, and [Vet1], [Vet2], [KSS] for some results on   hyperbolic billiards on a general Riemannian surface. There are many results in the literature concerning
hyperbolic dynamics for planar billiards
[Si], [Bu1-Bu4], [Wo2],  [Ma], [Do].  In  the present work we generalize
Wojtkowski's
criterion of hyperbolicity [Wo2] to  billiards on arbitrary surfaces of
constant curvature.

We interpret
Wojtkowski's condition [Wo2] in terms of a special class of trajectories,
which
generalize two-periodic orbits. Let $Q$ be a billiard table
on a surface of constant curvature.  The billiard  map $\phi :V\rightarrow
V$ acts
on the phase space $V$, which consists of pairs $v=(m,\theta)$.
Here $m$ is the position of the ball on the boundary $\partial Q$ of $Q$,
and $\theta$ is the angle between the outgoing velocity and
the tangent to $\partial Q$ at  $m$. The billiard  map
preserves a natural probability measure $\mu$ on $V$.
We denote the images of $v$  after
$n$ iterations by $(m_{n+1},\theta_{n+1}) =\phi^{n} (v)$.  The trajectory
$\phi^{n} (v)$
is a {\it generalized two-periodic trajectory } (g.t.p.t.) if the following
conditions are satisfied.\\
\\
\noindent 1. The incidence angle and the curvature of the boundary
$\kappa_{n}$
at the bouncing points have period 2:
$\theta_{2n}=\theta_{2}$, $\theta_{2n+1}=\theta_{1}$, $\kappa_{2n}=\kappa_2$,
$\kappa_{2n+1}=\kappa_1$;\\

\noindent  2. The geodesic distance between consecutive bouncing points
is constant: $s=|m_{n}m_{n+1}|$ (see fig. 1a). \\

\noindent If $\theta_i = \pi/2$,  the g.t.p.t. is a two-periodic orbit, see
fig. 1b.\\

Along a  g.t.p.t. the linearized map $D_{v}\phi$ is two-periodic, and the
stability of a  g.t.p.t. is determined by $D_{v}\phi^2$.
As we will see in  Section 2, for each surface of constant curvature,
the stability type of a g.t.p.t.  is
completely determined by the triple of parameters $(d_{1},d_{2},s)$, where
$2d_{1}$ (resp. $2d_{2}$) is the signed length of the chord generated by the
intersection of  the line $m_{1} m_{2}$ with the  osculating circle
at $m_{1}$ (resp. $m_{2}$) (see fig. 1a). We shall use the symbol
$T(d_{1},d_{2},s)$ for the g.t.p.t. with parameters $(d_{1},d_{2},s)$.

We will now discuss g.t.p.t.s for planar billiards in some detail.
Here $s$ is the euclidean distance between consecutive bouncing
points,  and  $d_{i}= r_{i}\sin\theta_{i}$, $i=1,2$,  where  $r_{i}$ are
the radii of  curvature of the boundary $\partial Q$ at the respective points.
If the curvature of the boundary at the bouncing point is zero we take $r_{i}=-\infty$ as the radius of  curvature and $d_{i}=-\infty$ respectively. By an elementary computation,  $T(d_{1},d_{2},s)$ is unstable if and only if

$$ s\in\cases{ [ d_{1},d_{2}] \cup  [ d_{1}+d_{2},\infty) &  if
$ d_{1},d_{2}\geq 0 $\cr
[0,\infty) &  if $d_{1},d_{2}\leq 0 $\cr
 [0, d_{1}+d_{2}]\cup [ d_{1},\infty) & if $d_{1}\geq 0,d_{2}\leq 0 $. \cr}
\eqno (1.1)
$$
Moreover, the trajectory is  hyperbolic (i. e., strictly unstable) if
$s$ is in the interior of the corresponding interval,
and the trajectory is parabolic if $s$ is a boundary point (in the limiting case $d_{1}=d_{2}=-\infty $ the trajectory is parabolic for any value of $s$).

We introduce the notions of B-unstable and S-unstable g.t.p.t.s.
The g.t.p.t. $T(d_{1},d_{2},s)$ is B-unstable if in eq. (1.1) $s$ belongs
to a ``big interval'':

$$ s\in\cases{[ d_{1}+d_{2},\infty) &  if $ d_{1},d_{2}\geq 0 $\cr
[0,\infty) &  if $d_{1},d_{2}\leq 0 $\cr
[ d_{1},\infty) & if $d_{1}\geq 0,d_{2}\leq 0 $.\cr} \eqno (1.2)
$$
On the contrary, if $s$ belongs to a ``small interval'', then
$T(d_{1},d_{2},s)$ is S-unstable:

$$ s\in\cases{[ d_{1},d_{2}] &  if $ d_{1},d_{2}\geq 0 $\cr
[ 0, d_{1}+d_{2}] & if $d_{1}\geq 0,d_{2}\leq 0 $.\cr} \eqno (1.3)
$$
Note that a small interval shrinks to a point when $|d_1|=|d_2|$.

We will outline a simple connection between the  present approach
and Wojtkowski's method (for planar billiards).
With any point $v=(m_1,\theta_1)\in V$ of the phase space
we associate a {\it formal} g.t.p.t. $T(v)$. Let
$\phi(v)=(m_{2},\theta_{2})$. We set $d_1=d(v)$, $d_2=d(\phi(v))$ and
$s=|m_{1} m_{2}|$.
The formal g.t.p.t. $T(v)$ can be realized as an actual g.t.p.t.
$T(d_{1},d_{2},s)$
in an auxiliary billiard table $Q_{v}$, constructed from the
boundary $\partial Q$ around $m_{i}$, as shown in fig. 2.

\vskip .2 cm

\noindent {\bf Definition 1.} {\it Let the notation be as above.
A point $v\in V$ of the billiard phase space is

a) B-hyperbolic (or strictly B-unstable) if the g.t.p.t. $T(v)$
is strictly B-unstable;

b)  B-parabolic if  $T(v)$ is B-unstable and parabolic
(i. e., $s$ belongs to the boundary of the appropriate big interval in
eq.(1.2));

c)  B-unstable if $T(v)$ is  B-unstable (i. e., B-parabolic or B-hyperbolic);

d)  eventually strictly B-unstable if for some $n \ge 0$ the point
$\phi^{n} (v)$ is strictly B-unstable, while $\phi^{i} (v)$ are
B-unstable for $0\le i <n$. } \\
\\

In our interpretation, Wojtkowski's hyperbolicity criterion [Wo2]
is the condition that $\mu$-almost all points  of the billiard phase space
are eventually strictly B-unstable.\\

The concept of g.t.p.t.s and the associated structures make sense for
billiards on any surface. In the body of the paper we will generalize the
notions
of the B-unstable and S-unstable g.t.p.t.s
to arbitrary surfaces of constant curvature, thus extending Definition 1 to
billiards
on all of these surfaces. Now we formulate the main
result of this work.\\
\\

\noindent {\bf  Theorem 1 (Main Theorem).} {\it Let $Q$ be a billiard table
on a surface of constant curvature, and let $\phi :V\rightarrow V$ be the
billiard  map. Let $\mu$ be the canonical invariant measure on $V$.
If $\mu$ almost every point of $V$ is eventually strictly B-unstable
then the billiard in $Q$ is hyperbolic.}\\
\\

Later on in the paper we will derive
geometric conditions on the billiard table that insure that  the g.t.p.t.s
are B-unstable. With these conditions, which depend on the
curvature of the surface, Theorem 1 will become a geometric criterion
for hyperbolicity of the billiard dynamics on surfaces of constant curvature.
In particular,  for planar billiards Theorem 1 yields Wojtkowski's
criterion [Wo2].

Let $\lambda(v) \ge 0$ be the Lyapunov exponent of the billiard
in the table $Q$, which is defined for $\mu$-almost all $v \in V$.
Recall that in our terminology the billiard in $Q$ is hyperbolic if
$\lambda(v)$
is positive $\mu$ almost everywhere. We denote by $h(Q)$ the metric entropy
(with respect to $\mu$) of the billiard in $Q$. Following the approach
of Wojtkowski's [Wo2], we will estimate from below the metric entropy
of billiards satisfying the conditions of Theorem 1.\\

Let $\phi_v$ be the map corresponding to the g.t.p.t.
$T(v)$, and let $\bar{\lambda}(v)=\lim_{n\to\pm\infty}\frac{1}{n}\log||D\phi^{n}_v||\ge 0$
be its Lyapunov exponent.\\
\\

\noindent {\bf  Theorem 2.} {\it Let $Q$ be a billiard table satisfying the
assumptions
of the main theorem, and let the notation be as above.
Then }

$$h(Q) \geq \int_{V}\bar{\lambda}(v)\,d\mu .\eqno (1.4)$$

To explain the mysterious appearance of g.t.p.t.s, which bear the crux of
our approach to
hyperbolicity in billiard dynamics, we will outline a connection between
them and
the method of invariant cone fields of Wojtkowski  [Wo1,Wo2]. Let $\sigma:V\rightarrow V$ be the time-reversal involution: $\sigma (m,\theta)= (m,\pi-\theta)$ and let ${\cal W}= \lbrace W(v) :  v \in V \rbrace $  be an invariant cone field defined in terms of a projective coordinate (each $W(v)$  is an interval in $\R\cup\infty$). We say that $\W$ is {\it  symmetric}, if $W(v)=W(\sigma(v))$ for each $v\in V$. The invariant cone fields defined in [Wo2] are symmetric.
It can be shown that the existence of a symmetric invariant cone field in $V$
implies
the instability of $\mu$ almost all g.t.p.t.s $T(v),\,v\in V$.  In the
proof of Theorem 1
we will show that for our class of billiards the (quasi)converse holds.
More precisely, if $\mu$ almost all g.t.p.t.s $T(v),\,v\in V,$ are
B-unstable, then  $V$
has a symmetric invariant cone field.  If, besides, $\mu$ almost all g.t.p.t.s
are eventually strictly B-unstable, then such cone field is eventually strictly invariant and the billiard dynamics is hyperbolic.

The plan of the paper is as follows. In Section 2 we provide the necessary
preliminaries and study the geometric optics (i. e., the propagation and
reflection
of infinitesimal light beams) on surfaces of constant curvature.
In Section 3 we apply these results to obtain explicit analogs of
eqs. (1.1-1.3).
We derive linear instability conditions
for g.t.p.t.s  and show that they  distinguish  between   B-unstable and
S-unstable trajectories in a natural way. In Section 4,
using invariant cone fields \'{a} l\'{a} Wojtkowski, we prove
the main theorem. We define our cone fields for billiards on all
surfaces of constant curvature. Employing  geometric optics,
we show that under the assumptions of the main theorem these cone fields
are invariant, and eventually strictly invariant.
Also in Section 4 we prove Theorem 2. In  Section 5 we derive
hyperbolicity criteria for elementary billiard
tables  (the boundary consists of circular arcs).
Then we apply the main theorem and its corollaries
to construct several classes of  billiard tables with hyperbolic dynamics
on the sphere and on the hyperbolic plane.
Finally, we formulate general principles for the design of billiard
tables satisfying the conditions of Theorem 1. In
particular,  we obtain the counterparts of Wojtkowski's
geometric inequality [Wo2]  for surfaces of constant nonzero curvature.
The calculations are involved, and we relegate them to the Appendix.

In  a forthcoming publication [Gb] we will apply the methods developed here
to investigate the dynamics of
billiards in constant magnetic fields on arbitrary surfaces of constant
curvature.

The results of  Wojtkowski [Wo2] have been strengthened (for
planar billiards) in [Bu3,Bu4], and [Do]. It turns out that the criteria
of [Bu3,Bu4],  and [Do]  can be obtained using certain
invariant cone fields, which are, in general, not symmetric. This suggests that
our hyperbolicity criterion for billiards on surfaces of constant curvature
can be considerably strengthened, by employing other invariant cone fields.
In particular, we believe that the results of Bunimovich [Bu3,Bu4] and
Donnay [Do]
can be extended to  billiards on surfaces of constant curvature.


\section{Geometric optics and billiards on surfaces of constant curvature}

Let $M$ be a simply connected surface of constant curvature, and
let $Q$ be a connected domain in $M$, with a piecewise smooth boundary
$\partial Q$.
For concreteness, we will assume that the curvature is either zero ($M=\R^2$),
or one ($M=\S^2$), or minus one ($M=\H^2$).
In what follows, $\partial Q$ is endowed with the positive orientation.

The billiard in $Q$ is the dynamical system arising from the geodesic
motion of a point mass inside $Q$, with specular reflections at the
boundary. The standard cross-section, $V\subset TQ$, of the billiard flow
consists of unit tangent vectors, with footpoints on $\partial Q$,
pointing inside $Q$. The first return associated with this cross-section
is the billiard map, $\phi:V\rightarrow V$.
We will use the standard coordinates $(l,\theta)$ on $V$,
where $l$ is the arclength parameter on $\partial Q$ and
$0\leq\theta\leq\pi$ is the angle between the vector and $\partial Q$.
We call $V$ the phase space of the billiard map, associated with the
billiard table $Q$. The invariant measure
$\mu=(2|\partial Q|)^{-1}\sin\theta dld\theta$ is a probability measure,
$\mu(V)=1$.

We will study the natural action of the differential of $\phi$ on the
projectivization
$B$ of the tangent manifold of $V$. Abstractly, $B$ consists of straight
lines (as opposed to vectors) in the tangent planes to points of $V$.
We will describe this space using the language of
geometric optics. An oriented curve $\gamma\subset M$, of class $C^2$, defines
a `light beam', i. e., the family of geodesic rays orthogonal to $\gamma$.
The geodesics which intersect $\gamma$ infinitesimally close to a point,
$m\in\gamma$, form an `infinitesimal beam', which is completely determined
by the normal unit vector $v\in T_mM$ to $\gamma$, and by the geodesic
curvature $\chi$ of $\gamma$ at  $m$. We denote the infinitesimal beam
by $b(v,\chi)$. Our convention for the sign of the curvature is opposite to
the
one used in [Si],  [Bu1-Bu4].

Infinitesimal beams yield a geometric representation of
the projectivized tangent manifold to the unit tangent bundle of $M$. In
particular, they give us a geometric realization of the space $B$.
We will describe the differential of the billiard map in this realization.
Let $p:B\rightarrow V$ be the natural projection.
Since $\dim V = 2$, each fiber $p^{-1}(v)\equiv B_v  \subset B$ is abstractly isomorphic
to the projective line, and we take $\chi\in\R\cup\infty$ as projective coordinate on $B_v$ (this  representation of $B$ was discussed for the planar case by e. g., [Wo2]).
In this coordinatization, $B_v= \{b(v,\chi):\chi\in\R\cup\infty\}$.

Let $X\subset TM$ be the set of unit tangent vectors with footpoints in
$\partial Q$,
and let $Y=\{b(v,\chi):v\in X,\chi\in\R\cup\infty\}$ be the set of
corresponding infinitesimal beams.
Let $\rho_m:T_mM\rightarrow T_mM$ be the linear reflection about the
tangent line to
$\partial Q$. As $m$ runs through $\partial Q$, the reflections $\rho_m$
yield a selfmapping
$\rho:X\rightarrow X$ whose differential acts on $Y$.

Let $\Phi^s$ denote the geodesic flow of $M$. Let $G(v)$ be the oriented
geodesic
defined by a unit tangent vector. For $v\in V$ let $s(v)$ be the distance along
$G(v)$ between the footpoint of $v$, and the next intersection point of $G(v)$
with $\partial Q$. Then $\Phi^{s(v)}(v)\in X$, and
$\rho\circ\Phi^{s(v)}(v)\in V$.
Let $\Phi:V\rightarrow X$ be the mapping $v\mapsto\Phi^{s(v)}(v)$.

We will use the same letters, $\phi$, $\rho$, and $\Phi$, for the
(projectivized) differentials
of these mappings. Since the billiard map is the composition:
$$ \phi = \rho\circ\Phi, \eqno(2.1)$$
it remains to compute the action of $\Phi$ and $\rho$ on infinitesimal beams.

Let $b(v_-,\chi_-)\in Y$ be an infinitesimal beam, and let $m\in \partial Q$
be the footpoint of $v_-$. Set $\rho\cdot b(v_-,\chi_-)=b(v_+,\chi_+)$.
Let $\kappa$ be the curvature of $\partial Q$
at $m$, and let $\theta$ be the angle between $v_-$ and the positive
tangent vector to
$\partial Q$ at $m$. Then $v_+=\rho_m(v_-)$, and

$$ \chi_+=\chi_- +{2\kappa\over\sin\theta}.\eqno(2.2)$$
This formula is well known when $M=\R^2$ [Si],  [Bu1], and extends
to all surfaces of constant curvature.

Let now $b=b(v,\chi)$ be an arbitrary infinitesimal beam, and set
$b'= \Phi^s\cdot b = b(v',\chi')$, where $v'\ = \Phi^s(v)$. We will express
$\chi'$ separately for each surface. \\
a) Flat case ($M=\R^2$).
By elementary euclidean geometry, we have
$$\chi'= {\chi\over 1 -s\chi }=-s^{-1}+{s^{-2}\over s^{-1}-\chi}. \eqno (2.3)$$
b) Curvature one case ($M=\S^2$).
By elementary spherical geometry:
$$
\chi'= -\cot s+{\sin^{-2}s \over\cot s-\chi}.\eqno(2.4)$$
c)  Curvature minus one case ($M=\H^2$).
The considerations depend on whether $|\chi|$ is greater or
less than one. However, the final expression is the same (we omit the details):
$$\chi'=-\coth s+{\sinh^{-2}s\over\coth s-\chi}.\eqno(2.5)$$

Note that in the limit $s\rightarrow 0$ eqs. (2.3-2.5) coincide.
For $v\in V$ set $D(v)=\sin\theta/\kappa$, so that eq. (2.2) becomes
$$-\chi_- + \chi_+={2\over D(v)}.\eqno(2.6)$$
Using classical formulas for surfaces of constant curvature
([Vi], compare also eq. (2.8) below with [Ta], for a different
but related context), we will give a geometric interpretation of the
function $D(\cdot)$.
Let $v\in V$, and let $m=m(l)\in\partial Q$ be the footpoint of $v$. Let
$C(l)\subset M$ be the osculating circle (hypercycle if $M=\H^2$ and
$|\kappa(l)| < 1$)
of $\partial Q$. The geodesic, $G(v)$, corresponding to $v$ intersects
$C(l)$ at $m$ and another
point, $m'=m(l')$. Let $\tilde{d}(v)$ be one half of the signed distance
between
$m$ and $m'$,
along $G(v)$. If $|\kappa(l)| < 1$, the hypercycle $C(l)$ consists of two
components, see fig. 3.
Then there are two possibilities: the points $l$ and $l'$ belong
to the same component (resp. different components) of $C(l)$, fig. 3. The
former
(resp. the latter) case occurs if $|D(v)|\leq 1$ (resp. $|D(v)|>1$).

\noindent{\it Remark:} When $\kappa(l)=0$ ($D(v)=\infty$) and $M=\R^2, \S^2$ there is  ambiguity  in the above definition of $\tilde{d}(v)$. In this case there are  two different values $\tilde{d}(v)=\pm \tilde{d}_{0}$ ($\tilde{d}_{0}=+\infty$ for  $M=\R^2$ and $\tilde{d}_{0}=\pi /2$ for  $M=\S^2$)  satisfying the above definiton (if $M=\H^2$, then $\tilde{d}_{0}=0$ and two values coincide). In what follows  we always choose in such case the negative value $-\tilde{d}_{0}$ as the definition for $\tilde{d}(v)$, i.e., we consider the case of zero curvature boundary as a limiting case of a  negative curvature boundary. Thus $\tilde{d}(v)\in[-\infty,\infty)$ if $M=\R^2$ and $\tilde{d}(v)\in[-\pi/2,\pi/2)$ if $M=\S^2$.

 Set
$$
d(v) = \cases {
\tilde{d}(v)& if $M=\R^2$ or $M=\S^2$ \cr
\tilde{d}(v)& if $M=\H^2$ and $|\kappa(l)| \ge 1$\cr
\tilde{d}(v)& if $M=\H^2,|\kappa(l)|<1$, $|D(v)|\leq 1$\cr
\tilde{d}(v)+i\pi/2& if $M=\H^2,|\kappa(l)|<1$, $|D(v)|>1$.\cr
}\eqno(2.7)
$$
Then we have
$$ D(v)=\cases{
d(v)& if $M=\R^2$ \cr
\tan d(v)& if $M=\S^2$ \cr
\tanh d(v)& if $M=\H^2$. \cr
} \eqno(2.8)
$$

For the case $M=\H^2$ we will use the following classification of points
of the phase space $V$. We say that $v\in V$ is of type $A$ (resp. $B$) if
$|D(v)|\le 1$ (resp. $|D(v)| > 1$). Let $V^A, V^B$ be the corresponding
subsets of $V$. Then $V=V^A\cup V^B$ is a partition. We will use the notation:
$$
\tilde{d}(v) = \cases{
d^A(v)\in[-\infty,\infty]& if $v\in V^A$ \cr
d^B(v)\in(-\infty,\infty)& if $v\in V^B$. \cr
}\eqno(2.9)
$$


\section{Generalized Two-Periodic Trajectories (g.t.p.t.s)}

Consider  the billiard dynamics in an arbitrary table  on a surface
of constant curvature. Eqs (2.2) and (2.3-2.5) describe the action of the
billiard
map on infinitesimal beams. Starting with an arbitrary $b(v,\chi)$ and
iterating the equations, we obtain for $\chi$ after infinite number of reflections a formal continued fraction

$$
c\equiv\chi^{\infty}=a_{0}+
{b_{0}\over\displaystyle a_{1}+
   { \strut b_{1} \over\displaystyle
a_{2}\cdots }} , \eqno(3.1)
$$
whose coefficients are determined by
$d_i=d(\phi^{i-1}\cdot v)$, and by the lengths $s_i$  of consecutive
billiard segments, where  $i = 1, 2, \dots$. The idea to associate a
continued fraction (3.1) to a billiard orbit has been introduced by Y. Sinai
in the seminal paper [Si], where he considered billiards in $\R^2$.
Eq. (3.1) is a direct extension of Sinai's idea to an arbitrary surface of
constant curvature.

Let $Q$ be a billiard table, and let $v\in V$ be an arbitrary point in the
phase space of the billiard map.
Set $v_1=v, v_2=\phi(v), d_i=d(v_i), i=1,2$,
and let $s=s(v)$ be the distance between the footpoints of
$v_1$ and $v_2$, respectively (fig. 2). Let $T(v) = T(d_1,d_2,s)$ be the
associated
g.t.p.t. (see Section 1). The g.t.p.t. $T(v)$  can be
realized as a trajectory in an artificial billiard table whose exact shape
$Q_v$ is not important (see fig. 2). We denote by $\phi_{v}$ the associated
billiard map.

 Let $c(v)$ be the formal continued fraction eq. (3.1),
corresponding to $T(v)$. Note that $c(v)$ is periodic. Proposition 1 below
relates the
convergence of $c(v)$ with the stability type of $T(v)$. Recall that the
standard
definitions of elliptic, hyperbolic, and parabolic periodic points can be
expressed
in terms of the appropriate power of the differential of the
transformation, i. e.,
a particular matrix associated with the periodic orbit, see, e. g., [KH].
Hence, these definitions straightforwardly extend to generalized periodic
orbits,
and we leave the details to the reader. In what follows we will talk about
elliptic,
parabolic, or hyperbolic g.t.p.t.s. We say that a g.t.p.t. is
(exponentially) unstable
if it is either hyperbolic or parabolic (resp. hyperbolic).

\noindent{\bf Proposition 1}. {\it Let $v\in V$ be arbitrary, and let the
notation be as
above.\\
The g.t.p.t. $T(v)$ is (exponentially) unstable if and only if the
continued fraction
$c(v)$ converges (exponentially fast).}

We outline a proof of Proposition 1, referring to [Wa] for the standard
material on continued
fractions. With a periodic continued fraction
one associates a fractional linear transformation, or, equivalently, a
$2\times 2$ matrix,
defined up to a scalar factor. For a $c(v)$ this matrix essentially coincides
with the linear transformation associated with the g.t.p.t. $T(v)$. The
claim now follows
from the standard facts [Wa] (we leave details to the reader).

Note that Proposition 1 (and its proof) straightforwardly extends to
generalized periodic trajectories of any period.

\noindent{\it Remark: }
Another approach to the stability of $T(v)$
is to consider the linearization $D\phi^{2}_{v}$. Then
$T(v)$ is
hyperbolic if $|tr(D\phi^{2}_{v})|> 2$, parabolic if $|tr(D\phi^{2}_{v})| =
2$, and elliptic if
$|tr(D\phi^{2}_{v})| < 2$.

\noindent{\bf Lemma 1}. {\it Let $v\in V$, and let $d_1,d_2,s$ be the
associated data.
Then the coefficients $a_i,b_i, i\ge 1$ of the continued fraction $c(v)$
are given
by the following formulas:

\noindent{a) $M=\R^2$. We have $a_{2n+1}=-2s^{-1}+2d^{-1}_{1}
,a_{2n}=-2s^{-1}+2d^{-1}_2, b_{n}=-s^{-2}$;}\\
\noindent{b) $M=\S^2$. Then $a_{2n+1}=-2\cot s
+2\cot d_{1}, a_{2n}=-2\cot
s+2\cot d_{2},
b_{n}=-\sin^{-2} s$;}\\
\noindent{c) $M=\H^2$. Here we have $a_{2n+1}=-2\coth s+2\coth d_{1},
a_{2n}=-2\coth s+2\coth d_{2}, b_{n}=-\sinh^{-2} s$.}}

\noindent{\it Proof}. The formulas are obtained by direct computations from
eqs. (2.2-2.6). \hfill$\Box$

Since the g.t.p.t. $T(v)$ and the continued fraction $c(v)$ are essentially
determined by the triple $(d_1,d_2,s)$ corresponding to $v$, we will use the
notation
$T(d_1,d_2,s)$ and $c(d_1,d_2,s)$ in what follows.
The formulas of Lemma 1 allow to compute the $2\times 2$ matrix associated with
$c(d_1,d_2,s)$. Analyzing this matrix for each of the three
surfaces, we obtain simple criteria for the convergence of $c(d_1,d_2,s)$.

\vspace{2mm}
\noindent{\bf Proposition 2.} {\it The continued fraction $c(d_1,d_2,s)$
converges
if and only if the following inequalities are satisfied.\\

\noindent{a)} If $M=\R^2$:
$$
(s-d_{1})(s-d_{2})(s- d_{1}-d_{2})s\geq 0. \eqno(3.2)$$

\noindent{b)} If $M=\S^2$:
$$
\sin (s-d_{1})\sin (s-d_{2})\sin(s-d_{1}-d_{2})\sin s \geq 0.\eqno(3.3)
$$

\noindent{c)} If $M=\H^2$:

$$
\sinh (s-d_{1})\sinh (s-d_{2})\sinh (s-d_{1}-d_{2})\sinh s \geq 0.\eqno(3.4)
$$
}

Taking into consideration that $s\geq 0$ for $\R^2$ and $\H^2$, and that
$0 \leq s \leq 2\pi$ for $\S^2$, we reformulate Proposition 2 in  a more explicit form.

\noindent a) Let $M=\R^2$. Then $T(d_1,d_2,s)$ is unstable if and only if

$$ s\in\cases{ [
d_{1},d_{2}] \cup  [ d_{1}+d_{2},\infty) &  if $ d_{1},d_{2}\geq 0
$\cr
[0,\infty) &  if $d_{1},d_{2}\leq 0 $\cr
 [0, d_{1}+d_{2}]\cup [
d_{1},\infty) & if $d_{1}\geq 0,d_{2}\leq 0 .$\cr} \eqno(3.5)
$$

\noindent b) Let $M=\S^2$. Set
$$
s\,\mod\pi = \cases{ s &  if $ s\le \pi $\cr
s-\pi &  if $ s > \pi. $\cr}
$$
Then  $T(d_1,d_2,s)$ is unstable if and only if
$$ s\,
\mod\pi\in\cases{ [d_{1}+d_{2},\pi]\cup [d_{1},d_{2}] &  if $
d_{1},d_{2}\geq 0 $\cr
[0,d_{1}+d_{2}+\pi]\cup [\pi-d_{1},\pi-d_{2}] &  if
$d_{1},d_{2}\leq 0 $\cr
[d_{2},\pi+d_{1}]\cup [0,d_{1}+d_{2}] & if
$d_{1}\leq 0 , d_{2}\geq 0, |d_{2}|\geq|d_{1}|$\cr
[d_{2},\pi+d_{1}]\cup
[\pi+d_{2}+d_{1},\pi]   & if $d_{1}\leq 0 , d_{2}\geq 0,
|d_{2}|\leq|d_{1}|.$\cr}
\eqno(3.6)
$$

\noindent c) Let $M=\H^2$.
We say that  $T(d_1,d_2,s)$ is of type $(A-A)$ if $v_1\in V^A$ and $v_2\in
V^A$.
The other types: $(A-B)$, $(B-A)$, and $(B-B)$ are defined analogously.
We formulate the criteria of instability for $T(d_1,d_2,s)$
`type-by-type'.

\noindent Type $(A-A)$:

$$
s\in\cases{ [ d_{1}^{A},d_{2}^{A}] \cup  [ d_{1}^{A}+d_{2}^{A},\infty) &
if $ d_{1}^{A},d_{2}^{A}\geq 0 $\cr
[0,\infty) &  if
$d_{1}^{A},d_{2}^{A}\leq 0 $\cr
 [0, d_{1}^{A}+d_{2}^{A}]\cup [
d_{1}^{A},\infty) & if $d_{1}^{A}\geq 0,d_{2}^{A}\leq 0 $.\cr} \eqno
(3.7a)
$$
\noindent Type $(B-B)$:

$$ s\in\cases{ [ d_{1}^{B}+d_{2}^{B},\infty)&
if $d_{1}^{B}+d_{2}^{B} \geq 0 $\cr
[0,\infty) & if $d_{1}^{B}+d_{2}^{B}\leq 0
$.\cr} \eqno (3.7b)
$$
\noindent Types $(A-B)$ or $(B-A)$:

$$
s\in\cases{ [
d_{1}^{A},\infty) &  if $d_{1}^{A}\geq 0 $\cr
[0,\infty) & if $d_{1}^{A}\leq
0$.\cr} \eqno(3.7c)
$$
It is worth mentioning that in eqs. (3.2-3.4) (resp. eqs. (3.5-3.7))
the hyperbolicity of $T(d_1,d_2,s)$ corresponds to strict inequalities (resp.
inclusions in the interior). The equality case (resp. boundary case)
corresponds to
the parabolicity of $T(d_1,d_2,s)$. There are also two special  cases when $T(d_1,d_2,s)$ is parabolic independently of the value of $s$: $M=\R^2$, $d_1=d_2=-\infty$ and $M=\H^2$, $|d_1|=|d_2|=\infty$ (it means also that $v_1, v_2\in V^A$).

We say that the right hand side in eqs. (3.5-3.7) is the instability set of
$T(d_1,d_2,s)$.
In general, it is a union of two intervals, where
one of them  degenerates  when $|d_1|=|d_2|$,
while the other is always nontrivial.
For want of a better name, we will say that the interval which persists is the
``big interval", and the other one is the ``small interval".
This motivates the following terminology: We will say that
$T(d_1,d_2,s)$
is (strictly) B-unstable if $s$ belongs to the (interior of the) big
interval of instability.
The proposition below makes this terminology explicit.

\noindent{\bf Proposition 3.}  {\it The g.t.p.t. $T(d_1,d_2,s)$ is
B-unstable if (and
only if) the triple $(d_1,d_2,s)$ satisfies the following conditions:}

\noindent{a)} {\it Let $M=\R^2$. Then }

$$
 s\in\cases{  [ d_{1}+d_{2},\infty) &  if $ d_{1},d_{2}\geq 0 $\cr
 [0,\infty) &  if $ d_{1},d_{2}\leq 0 $\cr
  [ d_{1},\infty) & if $d_{1}\geq 0,d_{2}\leq 0$. \cr} \eqno (3.8)
$$

\noindent{b)} {\it Let $M=\S^2$. Then }

$$ s\,
\mod\pi\in\cases{ [d_{1}+d_{2},\pi] &  if $ d_{1},d_{2}\geq 0
$\cr
[0,d_{1}+d_{2}+\pi] &  if $d_{1},d_{2}\leq 0 $\cr
[d_{2},\pi+d_{1}]
& if $d_{1}\leq 0 , d_{2}\geq 0$.\cr} \eqno(3.9)
$$

\noindent{c)}  {\it Let $M=\H^2$. Then: }

\noindent {\it In the case $(A-A)$ }

$$ s\in\cases{  [ d_{1}^{A}+d_{2}^{A},\infty) &  if $
d_{1}^{A},d_{2}^{A}\geq 0 $\cr
[0,\infty) &  if
$d_{1}^{A},d_{2}^{A}\leq 0 $\cr
 [ d_{1}^{A},\infty) & if $d_{1}^{A}\geq
0,d_{2}^{A}\leq 0 $, \cr} \eqno
 (3.10a)
$$
{\it or $|d_{1}^{A}|=|d_{2}^{A}|=\infty$ and arbitrary $s$.}

\noindent  {\it In the case $(B-B)$ }

$$
s\in\cases{ [ d_{1}^{B}+d_{2}^{B},\infty)&  if $d_{1}^{B}+d_{2}^{B} \geq 0
$\cr
[0,\infty) & if $d_{1}^{B}+d_{2}^{B}\leq 0 $.\cr} \eqno
(3.10b)
$$
\noindent {\it In the cases $(A-B)$ or $(B-A)$ }

$$
s\in\cases{ [ d_{1}^{A},\infty) &  if
$d_{1}^{A}\geq 0 $\cr
[0,\infty) & if $d_{1}^{A}\leq 0$.\cr}
\eqno(3.10c)
$$


\section{ Proofs of  Theorem 1 and Theorem 2}

{\it Proof of the main theorem (Theorem 1).} We will define a cone field on
the phase
space of the billiard  map.
A cone in $T_vV$ corresponds to an interval in the projectivization,
$B_v$. In Section 2 we have explicitly identified each space $B_v$
with the standard projective line $\R\cup\infty$. Therefore, a cone field,
$\W$, is determined by a function, $W(\cdot)$, on $V$, where each
$W(v)\subset\R\cup\infty$ is an interval in the projective coordinate $\chi$.

We introduce an auxiliary coordinate $f$ on $B_v$, which has a simple
geometric meaning. Let $b(v,\chi)$ be an infinitesimal beam,
and let $G(v)$ be the corresponding
oriented geodesic. Consider the beams $\Phi^t\cdot b(v,\chi)$, obtained
by the action of the geodesic flow.
Suppose, that $M=\R^2$ or $M=\S^2$, or $M=\H^2$ and $|\chi|\ge 1$.
Then there is $t\in \R\cup\infty$, such that
the beam $\Phi^t\cdot b(v,\chi)$ has infinite
curvature. If $M=\R^2$ or $M=\H^2$ ($|\chi|\ge 1$), then $t$ is unique, and we set $f(\chi)=t$.
If $M=\S^2$,
then $t$ is unique modulo $\pi$, and let $f(\chi)\in [-\pi/2,\pi/2)$ be the one
with the smallest absolute value. We denote by $o(v,\chi)\in M$ the
footpoint of $\Phi^{f(\chi)}\cdot v$. This is the focusing point of the
infinitesimal beam $ b(v,\chi)$, see fig. 4a,b,c.
If $M=\H^2$, and  $|\chi|<1$ then the beam $b(v,\chi)$ has no
focusing point fig. 4d.

While the focusing point, $o(v,\chi)$, depends on both $v$ and
$\chi$, the signed focusing distance is determined
by the curvature of the beam alone, $f=f(\chi)$. The explicit relations
between $f$ and $\chi$ depend on $M$.\\
a) When $M=\R^2$, we have $\chi=1/f$;
b) If $M=\S^2$, we have $\chi= \cot(f)$;
c) If $M=\H^2$ and $|\chi|\ge 1$, we have $\chi=\coth(f)$.

We will define the cone field $\W$ using the projective coordinate $\chi$.

\noindent a)  Let $M=\R^2$. Set

$$
W(v)=\cases{ [-\infty, D^{-1}(v)]&  if $D(v)\leq 0 $\cr
 [D^{-1}(v), +\infty]& if $D(v)> 0 $.\cr}
$$

\noindent b) Let $M=\S^2$. Set

$$ W(v)=\cases{ [-\infty, D^{-1}(v)] &  if $D(v)\leq 0$\cr
 [D^{-1}(v), +\infty]& if $D(v)> 0 $.\cr}
$$

\noindent c) Let $M=\H^2$. We consider two cases.

\noindent 1) If $v\in V^A$, we set

$$
W(v)=\cases{ [-\infty, D^{-1}(v)] &  if $D(v)\leq 0$\cr
[D^{-1}(v), +\infty] & if $D(v)>0 $.\cr}
$$

\noindent 2) If $v\in V^B$, then

$$
W(v)=[-\infty, D^{-1}(v)].
$$

In terms of the auxiliary coordinate $f$ the cone field $\W$ is
given  for $M=\R^2$ and  $M=\S^2$ by the following intervals:

$$
 W(v)=\cases{ [d(v),0]&  if $d(v)\leq 0 $\cr
 [0,d(v)]& if $d(v)> 0 $.\cr}
$$
In what follows, we will use the cone field $\W$ in one form or the other,
whichever is more convenient.

We recall the classification of points in the phase space of the billiard
map. A  point $v\in V$ is B-hyperbolic (we will also say
strictly B-unstable) if the corresponding g.t.p.t. $T(v)$ is  B-unstable
and hyperbolic. A point is B-parabolic if $T(v)$ is  B-unstable and
parabolic. Putting the two definitions together, we will say that
$v\in V$ is B-unstable if the corresponding g.t.p.t.
$T(v)$ is  B-unstable (i. e., either B-parabolic or B-hyperbolic).
We will say that $v\in V$ is eventually strictly B-unstable
if there exists $n\ge 0$ such that the points
$\phi^i(v)$ are B-unstable for $0\leq i <n$ and $\phi^n(v)$ is strictly
B-unstable.


\noindent {\bf Lemma 2.}                
{\it Let $M$ be a surface of constant curvature, let $Q\subset M$ be an
arbitrary
billiard table, and let $\W$ be the cone field defined above. Let $v\in V$
be such that the g.t.p.t. $T(v)$ is (strictly) $B$-unstable. \\
Then $\phi(W(v)) \subseteq W(\phi(v))$ (resp. the strict inclusion
$\phi(W(v)) \subset W(\phi(v))$ holds).}

\noindent {\it Proof.} Let $(d_1,d_2,s)$ be the triple, associated to $v$.
We will prove the claim separately for each of the three surfaces.

\noindent a) Let $M=\R^2$ (fig. 5). We rewrite eq. (2.6) as
$$
 {(s-f_{1}-d_{2}) \over s- f_{1} }= {(d_{2}-f_{2}) \over f_{2}}.\eqno(4.1)
$$
Since $(d_1,d_2,s)$ satisfies eq. (3.8), we obtain
$(d_2-f_2)/ f_{2}\geq 0$. The inequality is strict if $T(v)$ is strictly
$B$-unstable. This implies the claim.

\noindent b) Let $M=\S^2$ (fig. 5). Eq. (2.6) and the relation between $\chi$ and
$f$ on $\S^2$
imply
$$
{\sin (s-f_{1}-d_{2})\over
\sin (s-f_{1})}={\sin (d_{2}-f_{2})\over \sin f_{2}}.\eqno (4.2)
$$
Since the triple $(d_1,d_2,s)$ satisfies eq. (3.9),
$\sin (d_{2}-f_2)/\sin f_{2}\geq 0$ (strict inequality if $T(v)$ is strictly
$B$-unstable). Simple considerations, which we leave to the reader, yield
the claim.

\noindent c) Let $M=\H^2$. From eqs. (2.5) and (2.6) we have

$$\chi_2 = \frac{2}{D(v)} - \coth s + {\sinh^{-2}s\over\coth
s-\chi_1}.\eqno (4.3)$$

\noindent Recall that $V=V^A\cup V^B$, a partition of $V$ into the sets of
points of type
$A$ and type $B$. Hence, depending on the type of $v_i,\, i=1,2$, we have
four cases
to consider. We will prove the claim case-by-case.

\noindent Case $B-B$. From eq. (4.3) and eq. (3.10b), we obtain $\chi_2 \le
\tanh d_2^B$,
which implies the claim.

\noindent Case $B-A$.  From eq. (4.3) and eq. (3.10c), we have $\chi_2 \in
[-\infty,\coth d_2^A]$
if $d_2^A\le 0$, and $\chi_2 \in [\coth d_2^A,\infty]$ if $d_2^A > 0$. The
claim
follows.

\noindent  Case $A-A$.  From eq. (4.3) and eq. (3.10a), we obtain $\chi_2
\in [-\infty,\coth d_2^A]$
if $d_2^A\le 0$, and $\chi_2 \in [\coth d_2^A,\infty]$ if $d_2^A > 0$, which
implies the claim.

\noindent Case $A-B$.  From eq. (4.3) and eq. (3.10c), we have $\chi_2 \le
\tanh d_2^B$,
implying the claim. This proves Lemma 2.  \hfill$\Box$


Now we finish the proof of the main theorem. Since, by
assumption, almost every point of the phase space is eventually
strictly B-unstable, Lemma 2 implies that the cone field $\W$ is
eventually strictly
invariant. The claim now follows from a theorem of Wojtkowski
[Wo1], [Wo2].   \hfill$\Box$


\noindent {\it Proof of  Theorem 2.}
Let $l(v)$ and $r(v)$ be the left and the right endpoints of the interval
$W(v)$
defined in terms of the projective coordinate (for  the cone fields
defined above $l(v)$ and $r(v)$ are either $\infty$ or $D^{-1}(v)$). Let
$l_{1}(v)$ and $r_{1}(v)$ be the left and the right endpoints of the
interval $\phi(W(v))$.
Applying Theorem 2 in [Wo2] to the billiards,
satisfying the assumptions of the main theorem, we obtain

$$
\int_{V}\lambda_{+}\,d\mu \geq
\int_{V}\log{\frac{\sqrt{\zeta}+1}{\sqrt{\zeta-1}} }
\,d\mu ,\eqno (4.4)
$$
where

$$
\zeta(v)=\frac{r(\phi(v))-l_{1}(v)}{r(\phi(v))-r_{1}(v)}
\, \, \frac{l(\phi(v))-r_{1}(v)}{l(\phi(v))-l_{1}(v)}
$$
Let $\phi_{v}$ be the map associated with the g.t.p.t. $T(v)$.
By straightforward calculations

$$\left( \frac{\sqrt{\zeta}+1}{\sqrt{\zeta-1}}  \right)^2
+ \left( \frac{\sqrt{\zeta}+1}{\sqrt{\zeta-1}}
\right)^{-2}=|tr(D\phi_{v}^{2})|.$$
The claim now follows from the inequality (4.4).   \hfill$\Box$


\section{ Applications and Examples }

  There are many classes of planar domains with hyperbolic billiard dynamics
 [Wo2], [Bu3,4], [Ma]; see also [Tab] and the references there.
 In subsection 5.1 we will apply the main Theorem to
obtain convenient sufficient conditions of hyperbolicity for elementary
billiard tables on
all surfaces of constant curvature.
In subsection 5.2 we will use these conditions (as well as
the main Theorem directly) to construct
several classes of examples of billiard tables with chaotic dynamics on
$\S^2$ and  $\H^2$. In subsection 5.3, expanding the ideas of [Wo2] for
billiards in $\R^2$, we
obtain a simple set of principles for constructing
billiard tables with hyperbolic dynamics on arbitrary surfaces of constant
curvature.

\subsection{Elementary billiard tables:
conditions for hyperbolicity}            


We shall use the term ''elementary billiard tables'' to denote billiard tables $Q$, such that $\partial Q$
is a finite union of arcs, $\Gamma_i$, of constant geodesic curvature,
$\kappa(\Gamma_i)=\kappa_i$. We will use the notation
$\Gamma_i^+$ (resp. $\Gamma_i^-$, resp. $\Gamma_i^0$) to
indicate that $\kappa_i>0$ (resp. $\kappa_i<0$, resp. $\kappa_i=0$).
Let $C_i$ be the curve of constant curvature containing $\Gamma_i$.
Let $D_i\subset M$ be the smallest region such that $C_i =\partial D_i$.
The representation $\partial Q=\cup_{i=1}^N\Gamma_i$
is unique, and we call $\Gamma_i$ the components. We will refer to
$\Gamma_i^+$ (resp. $\Gamma_i^-$, resp. $\Gamma_i^0$) as the components of type
plus (resp. of type minus, resp. of type zero).

Applying the main Theorem to elementary
billiard tables in $\R^2$, we recover a classical result of L. Bunimovich
[Bu1].

\noindent {\bf Corollary 1.} {\it Let $Q\subset\R^2$ be an elementary
billiard table
with at least two boundary components, and assume that not
all of them are neutral. If for every
$\Gamma_i^+$ we have $D_i\subset Q$, then the billiard in $Q$ is hyperbolic.}

The extension of this result for $M=\S^2$ and $M=\H^2$ will be given below. For this purpose we introduce the following terminology: If  
 $R\subset S\subset M$ are regions with piecewise $C^1$ boundaries, we call
the inclusion $R\subset S$  {\it proper} if $\partial R\cap
int\,S\ne\emptyset$.

Consider now an elementary billiard table $Q\subset \S^2$. For any domain $D\subset \S^2$ we denote by $-D\subset \S^2$ the domain obtained  by the reflection of $D$ about the center of the sphere (polar domain).

\noindent {\bf Condition S1}. The table $Q$ satisfies $D_i\subset Q$
for every boundary component $\Gamma_{i}^{+}$. Besides, either
$-D_i\subset Q$, or $-D_i\subset \S^2\setminus Q$, and the inclusions are
proper.

\noindent {\bf Condition S2}.  For every
$\Gamma_{j}^{-}$ we have $D_j\subset \S^2\setminus Q$, and the inclusions
$-D_j\subset \S^2\setminus Q$, or $-D_j\subset Q$ are proper.

\noindent {\bf Corollary 2.}\ {\it  Let $Q\subset \S^2$ be an elementary
billiard table with at least two boundary components of nonzero type.
If $Q$ satisfies conditions S1 and S2, then the billiard in $Q$ is
hyperbolic.}

\noindent  Outline of proof: Straightforward analysis
shows that $Q$ satisfies the conditions of the main Theorem.

\noindent{\it Remark}. Suppose $Q'=\S^2\setminus Q$ is connected. If $Q$
satisfies conditions S1 and S2, then $Q'$ also does, and hence the billiard
in $Q'$ is hyperbolic.

Let $Q\subset \H^2$ be an elementary billiard table. We use the
notation $\Gamma_i^A$ (resp. $\Gamma_i^B$) if $|\kappa_i|\ge 1$ (resp.
$|\kappa_i| < 1$).
In combination with the previous conventions, this yields the
self-explanatory notation
$\Gamma_i^{A+},\Gamma_i^{A-},\Gamma_i^{A0},\Gamma_i^{B+}$, etc.
We will call them the components of type A plus, B minus, etc.

\noindent {\bf Condition H1}. For every component $\Gamma_i^{A+}$ of
$\partial Q$,
we have $D_i\subset Q$.

\noindent {\bf Condition H2}. There are no components of type $B+$.

\noindent {\bf Corollary 3.} {\it Let $Q\subset \H^2$ be an elementary
billiard table with at least two boundary components. If $Q$ satisfies
conditions
H1 and H2, then the billiard in $Q$ is hyperbolic.}

\noindent Outline of proof: The assumptions of Corollary 3 imply those of
the main Theorem.

\noindent{\it Remark}. The purpose of the assumptions that $\partial Q$ has
at least two
boundary components, and that the inclusions are proper is to exclude
degenerate situations, where each $v\in V$ is B-parabolic.  For instance,
this is the case if $Q$ is a disc, or an annulus between concentric circles.


\subsection{ Elementary hyperbolic billiard tables:  examples} 

Using Corollaries 2 and 3, we will produce examples of elementary
billiard tables with hyperbolic dynamics in $\S^2$ and $\H^2$.
Besides, we will give examples of elementary billiard tables
that do not satisfy the assumptions of Corollaries 2 and 3, but have
hyperbolic dynamics. We will prove the hyperbolicity of these billiards
from the main Theorem.

\noindent 5.2a) {\bf Examples on the sphere}

\noindent{\it  Spherical Lorenz gas}.
One of the first examples of hyperbolic billiards was the flat torus with a
round hole, i. e., the Sinai billiard. This dynamical system is the
simplest special case of the Lorenz gas, which is still actively investigated.
The natural analog of the Lorenz gas on the sphere is the billiard table,
obtained by removing a finite number of disjoint discs, see fig. 6a.

Removing one disc, or a pair of parallel discs, we obtain an integrable
billiard
[Ve]. Let $D_i, 1 \le i \le n$, be the removed discs, so that
$Q=\S^2\setminus\cup D_i$, and $n>1$.
If all intersections $D_i\cap \pm D_j, i\ne j,$ are empty, then the
billiard in $Q$ is hyperbolic,
by Corollary 2, see fig. 6b for $n=2$. For these billiards the
non-intersection
condition above is also necessary for hyperbolicity.
If it is not satisfied, then $Q$ has stable periodic  orbits of period
two. They go
along the large circle which connects the centers of the two removed discs.

Let now $Q$ be obtained by removing $m$ pairs of parallel discs, $P_i, 1
\le i \le m$, and
$n$ single discs, $D_j, 1 \le j \le n$, where $m+n > 1$. Consider the
configuration
$(\cup_{i=1}^m\pm P_i)\cup(\cup_{j=1}^n\pm D_j)$. Suppose that the only
nonempty
intersections are the trivial ones: $P_i\cap-P_i\ne\emptyset$, see fig. 6c.
Corollary 2 does not apply, however a direct analysis shows that almost every
point of the phase space is eventually strictly B-unstable. By the main
Theorem, these
billiard tables are hyperbolic.

\noindent {\it Pseudo-stadia}. A pseudo-stadium on $\S^2$ is
an elementary billiard table $Q$, such that $\partial Q$ has four components :
Two of them are parallel, and of negative type, and the other two are of
positive type, see fig. 7. The two positive components may have the same
curvature, fig. 7a, or different curvatures, fig. 7b. If   $Q$ satisfies the
conditions of  the main
Theorem (like the pseudo-stadia in figs. 7a, 7b), then $Q$ is hyperbolic.

\noindent {\it Flowers}. Figs. 8a,b,c are examples of elementary billiard
tables, that belong to the class
of ``flowers". Some flowers satisfy the conditions of Corollary 2,
and hence, are hyperbolic.
Note that the dual tables $Q'=\S^2\setminus Q$
satisfy the conditions of Corollary 2 as well (see figs. 8a,b,c). Hence,
they are also hyperbolic,

\noindent {\it Billiard tables with flat components}.
Let a billiard table $Q\subset \S^2$ (not necessarily elementary) have a flat
component, $\Gamma^0\subset\partial Q$. We apply to $Q$ the method of
reflections, widely used to study billiards in polygons [Ge]. In a
nutshell, we associate with $Q$ the table $Q_1$, which is the union of $Q$
and its reflection about $\Gamma^0$, see fig. 9a. The
billiard dynamics in $Q$ and $Q_1$ are essentially isomorphic.
(We leave it to the reader to extend the argument of  [Ge] from $\R^2$ to all
surfaces of constant curvature.) Hence, if $Q_1$ satisfies the conditions of
the main
Theorem, then the billiard in $Q$ is hyperbolic.

Sometimes the method of reflections yields an easy proof of hyperbolicity.
Fig. 9a illustrates this point: The table $Q$ in fig. 9a
does not satisfy the conditions of Corollary 2, but $Q_1$ does. The preceding
discussion implies that $Q$ is hyperbolic.

Let $\partial Q$ have two or more flat components. Then, typically, $Q$
does not satisfy conditions of the main Theorem.  Let $Q_1$ be the table,
obtained by ``reflecting and unfolding'' $Q$ about the flat
components any number of times (including infinity). Often, $Q_1$ is not a
subset of $\S^2$
because of self-intersections. Then we think of $Q_1$ as a billiard table
located in a branched
covering of $\S^2$. Unfolding $Q$ infinitely many
times, we can always assume that
$Q_1$ has no flat components in its boundary. However, typically, the phase
space of $Q_1$ will have points $v$ such that in the corresponding triple
$(d_1,d_2,s)$ the distance $s$ is near $\pi$. Therefore, $Q_1$ does not
satisfy the conditions of the main Theorem. See, for example, the stadium
in fig. 9b.
If $Q_1$ is located strictly inside a hemisphere (possibly with
self-intersections), then this
problem does not arise. In particular, if $Q_1$ satisfies
the conditions of Corollary 2, then the billiard dynamics in $Q$
is hyperbolic. For instance, in figs. 9c and 9d, $Q_1$ is inside the upper
hemisphere, and satisfies the conditions of main Theorem. Hence, the ``stadia"
in figs. 9c and 9d have hyperbolic billiard dynamics.

\vspace{3mm}
\noindent 5.2b) {\bf Examples on the hyperbolic plane}

\noindent{\it Analogs of the Sinai billiard}.
Consider the billiard tables $Q\subset \H^2$ (not necessarily elementary)
such that $\partial Q$ has components
of nonpositive curvature only (fig. 10a). Let $v\in V$. If
$v\in V^A$, then
$d^A(v)\le 0$, and for $v\in V^B$ we also have $d^B(v)\le 0$. By eq. (3.10),
$Q$ satisfies conditions of main Theorem, hence these billiard
tables have hyperbolic dynamics.

\noindent{\it Polygons}.
Let $Q$ be a geodesic polygon in $\H^2$, see fig. 10b.
Then $V=V^B$, and $d^B (v)=0$ for every $v\in V$.
By eq. (3.10b), $Q$ satisfies the assumptions of main Theorem. Thus,
geodesic polygons
in $\H^2$ have  hyperbolic dynamics. In fact, polygons are a special case of
the Sinai billiards in $\H^2$.

\noindent{\it Stadia}. Let $Q\subset \H^2$ be an analog of the stadium:
$\partial Q$ has
four components, two of type zero, and two of positive type (fig. 11).
Let $Q$ be any stadium, and let $Q_1$ be the table
obtained by unfolding $Q$ about the flat components infinitely many times,
see fig. 11. If $Q_1$ satisfies the conditions of Corollary 3, then, applying
the method of reflections [Ge], extended to
the hyperbolic  plane, we obtain that the billiard in $Q$ is hyperbolic.
Figs. 11 illustrate this point.

\noindent {\it Flowers}. This is another class of elementary billiard
tables in $\H^2$ (fig. 12). If $\partial Q$ satisfies conditions H1 and H2
(see fig. 12a and 12b), then, by  Corollary 3, the billiard in $Q$ is
hyperbolic.


\subsection{Convex scattering for billiards on surfaces of constant
curvature}

Let $M$ be a surface of constant curvature.
In this subsection we consider billiard tables in $M$ with piecewise smooth
boundary, $\partial Q=\cup_{i}\gamma_i$. We will investigate the conditions on the components $\gamma_{i}$ which ensure that the billiard in $Q$ is hyperbolic.

In [Wo2] Wojtkowski introduced the notion of convex scattering.
By definition, a convex arc $\gamma\subset\R^2$ is convex scattering,
if it can be used as a component of a billiard table, for which the cone field
defined in [Wo2] is invariant. Using the notion of convex scattering,
Wojtkowski introduced
three ``principles of design of billiards (in $\R^2$) with hyperbolic
dynamics'', and constructed several examples of such tables.

  In our notation, $\gamma\subset\R^2$ is convex scattering if for any $v\in V$,
such that the footpoints of $v$ and $\phi(v)$ belong to $\gamma$ the corresponding g.t.p.t. $T(v)$ is B-unstable. Such condition is equivalent (see eq. 3.8) to the inequality
$d_1+d_2\le s$ as it appears in [Wo2]. Let $l$ be the arclength parameter on $\gamma$, and let $r(l)$ be the
radius of curvature. A convex arc $\gamma$ is convex scattering if and
only if  $r''\le 0$, as it has been shown in [Wo2]. In what follows we
generalize the notion of convex scattering to $\S^2$ and $\H^2$. We call a convex curve $\gamma\subset M$  convex scattering if for any $v\in V$,
such that the footpoints of $v$ and $\phi(v)$ belong to $\gamma$ the corresponding g.t.p.t. $T(v)$ is B-unstable. Using Proposition 3 we will obtain
 geometric criteria for convex scattering.
Then we will extend to $\S^2$ and $\H^2$ Wojtkowski's principles of design
of billiards with hyperbolic dynamics.

\vspace{3mm}
\noindent 5.3a)   {\bf Convex scattering and hyperbolic billiard tables in
$\S^2$}.

\noindent A convex curve $\gamma\subset \S^2$ is convex scattering if for every pair of the points $\gamma_0, \gamma_1\in\gamma$, such that the arc of $\gamma$ between $\gamma_0$ and $\gamma_1$ lies entirely on one side of the geodesic passing through $\gamma_0$ and $\gamma_1$, we have

$$d_1+d_2\leq s \leq \pi\eqno (5.1)$$

\noindent (compare with condition (3.9)). For simplicity of exposition, we
will restrict
our attention to piecewise convex billiard tables.
The main theorem yields the following principles for the design
of billiard tables in $\S^2$ with hyperbolic dynamics:

\noindent P1: All components of $\partial Q$ are convex scattering.

\noindent P2: Every component of $\partial Q$ is sufficiently far, but
not too far, from the other components.

More precisely, condition P2 means that any two
consecutive bouncing points of the billiard ball satisfy eq. (5.1), even if
they belong to
different components of the boundary. In particular,
the interior angles between consecutive components of $\partial Q$ are
greater than $\pi$.

Let $\kappa(l)$ be the geodesic curvature of $\gamma$. In Appendix A we will show that the differential inequality $(\kappa^{-1})''\leq 0$
is necessary, but, in general, not sufficient for convex scattering. However, a
sufficiently short arc satisfying $(\kappa^{-1})''< 0$  is convex scattering.

Let $S_a$ be the spherical analog  of the cardioid. It is the curve
obtained by rotating a circle of radius $a$ on another circle of the same
radius, see fig. 13.
For small $a$ the curve $S_a$ is well approximated by the cardioid $R_a$.
Since $R_a$ is (strictly) convex scattering [Wo2], the curvature,
$\kappa_a$, satisfies
the inequality $\lim_{a\rightarrow 0} (\kappa^{-1}_{a})''< 0$. Since
$\tan r \sim a$, $(\tan r)' \sim a^0$   and $(\tan r)'' \sim a^{-1}$, as
$a$ goes to zero, condition (A.5) is satisfied for sufficiently small
$a$. Thus, there is a critical value,
$a_{cr}$, such that for $a<a_{cr}$ the curve $S_a$ is convex scattering, and
the billiard in it is hyperbolic.  This approach
generalizes to any curve on the sphere whose planar counterpart is
strictly convex scattering.

Finaly, let us mention here, that the application of the main theorem to the concave billiards on the sphere leads to the hyperbolicity criterion, which is closely related to the results of Vetier [Vet1] [Vet2] (see also [KSS]). In fact, if concave billiard on the sphere satisfies  Vetier conditions (conditions 1.2-1.4 in [KSS]) it satisfies also the conditions of the main theorem.    

\vspace{3mm}
\noindent 5.3b)  {\bf  Convex scattering and hyperbolic billiard tables in
$\H^2$}.

\noindent A convex curve $\gamma\subset\H^2$ is convex scattering if for
each $v\in V$,
such that the footpoints of $v$ and $\phi(v)$ belong to $\gamma$, we have
$v,\phi(v)\in V^{A}$ and

$$d_1+d_2\leq s \eqno (5.2)$$

\noindent (compare with eq. (3.10)).
The differential inequality $(\kappa^{-1})''\leq 0$ is
necessary but, in general, not sufficient for eq. (5.2),
see Appendix B. $(\kappa^{-1})''< 0$ implies that every sufficiently
short arc is convex scattering.

The main theorem yields the following principles for the design
of billiard tables in $\H^2$ with hyperbolic dynamics:

\noindent P1: All convex components  of $\partial Q$ are  convex scattering.

\noindent P2: Every convex component of $\partial Q$ is sufficiently far from
any other component and  satisfies $\kappa (l)\geq 1$.

More precisely, condition P2 means that any two
consecutive bouncing points of the billiard ball which belong to different
components  satisfy eq. (3.10). This
implies the following conditions on the angles between adjacent components of
$\partial Q$.

\noindent P3: Let $\gamma',\gamma''\subset\partial Q$ be two adjacent
components.
If they are both convex, then the angle between them is greater than $\pi$.
If one of them is convex and the other is concave, then the angle
is greater  than or equal to $\pi$.

\noindent {\it Remark.} Comparing the principles of the design of hyperbolic
billiard tables for the three types of surfaces
of constant curvature, we see the same pattern.
There are, however, important differences. For instance, on $\S^2$, we need
to complement
the requirement ``to be far from each other'' for the components of
$\partial Q$,
by the one ``to be not too far''. The other important difference is that
on $\S^2$ and $\H^2$
the differential inequality $(\kappa^{-1})''\leq 0$ is necessary, but not
sufficient for convex scattering, see the Appendix below.


\section* { Acknowledgments}

 This work was supported partially by the Minerva Center for Nonlinear
Physics of Complex Systems.


\section { Appendix: geometry of convex scattering on $\S^2$ and $\H^2$}

We will investigate when a convex arc on the sphere or the hyperbolic plane
is convex scattering.

Let $M$ be any surface of constant curvature.
Let $\gamma\subset M$ be any smooth curve, and
let $\kappa(l)$ be the geodesic curvature of $\gamma$ (as a function of
arclength).
Let $r(l)$ be the radius of the osculating
circle (hypercycle if $M=\H^2$, and $|\kappa(l)|< 1$).
Then $\kappa=r^{-1}$ in $\R^{2}$, and $\kappa=\cot r$ for $\S^{2}$.
On $\H^2$ we will modify the definition of $r(l)$. There are two cases, A and B
(compare with section 2), where $|\kappa(l)| > 1$ in case A, and
$|\kappa(l)|\le 1$
in case B.  We will denote by $r^A$ and $r^B$ respectively the radius of
the osculating circle
(hypercycle).
In the case A (resp. B) we have $\kappa=\coth r^{A}$ (resp. $\kappa=\tanh
r^{B}$).
We set $r = r^{A}$ and $r=r^{B}+i\pi/2$ respectively. Then $\kappa=\coth r$.


\subsection* {A:  The sphere}

Let $\alpha$ and $\beta$ be a pair of orthogonal oriented geodesics on $\S^2$.
For $A\in \S^{2}$ let $x$ and $y$ be the oriented
distances from $A$ to $\alpha$ and $\beta$. Then
$(x,y)$ is a coordinate system in  $\S^2$.

Let  now $\gamma (l_{0})$ and $\gamma (l_{1})$ be two points on
$\gamma$ such that the arc of  $\gamma$ between $\gamma (l_{0})$ and
$\gamma (l_{1})$
lies on one side of the geodesic passing through these points, see fig.
14a. Let
$\alpha$ be that geodesic, and let $\beta$ be such that in the parameterization
$\gamma(l)=(x(l),y(l)),\ l_0 < l < l_1,$
the coordinate $y$ takes its maximal value when $x=0$, see fig. 14a. Let
$\theta (l)$ be the angle between $\gamma$ and the orthogonal to $\beta$
geodesic
passing through $\gamma(l)$. By elementary geometry:

$$
 {dx\over dl}=\cos\theta,  \qquad  {dy\over dl}={\sin\theta \over \cos
x},\eqno (A.1a)
$$
 $$             {d\theta\over dl}=\sin\theta \tan x -\cot r   .\eqno (A.1b) $$
Since $\gamma$ is convex, the inequality $s< \pi$ in eq. (5.1) is satisfied
for any two points of $\gamma$.

It remains to consider the inequality $s\geq d_{1}+d_{2}$.
Set $\Delta =s-d_{1}-d_{2}$. Then

\begin{eqnarray*}
 \Delta &=&\int [d(\arctan(\tan r\sin\theta ))+dx]\\
        &=&  \int dy{ \Big( (\tan r )'+ \cos\theta\tan r(\tan
x+\sin\theta\tan r )
\Big)\cos x \over 1+\tan^{2} r \sin^{2}\theta}.
\end{eqnarray*}
Since $y(l_{0})=y(l_{1})=0$, we obtain

$$\Delta =-\int dl\, \Big( (\tan r )''+F(\theta(l),r(l),x(l)) \Big)
{y\cos x \over 1+\tan^{2} r \sin^{2}\theta} \eqno (A.2) $$
where we have set for brevity

\begin{eqnarray*}
\lefteqn{F(\theta,r,x)}\\
 &=&\tan r\sin^{2}\theta (1-\tan^{2} x)-\sin^{3}\theta\tan x\tan^{2}r
+\sin\theta\tan x \\
 &+&(\tan r)'\tan r\sin 2\theta -{(\tan r )'+ \cos\theta\tan r(\tan
x+\sin\theta\tan r ) \over
1+\tan^{2} r \sin^{2}\theta}  \\
 &\times&( (\tan^{2} r )'\sin^{2}\theta +\tan^{2} r
\sin\theta\sin2\theta\tan x-\tan r\sin 2\theta)
\end{eqnarray*}
Set $L=l_1-l_0$. From eq. (A.2) we have

$$
\Delta  = -{(\tan r)'' \over
12\tan r}L^{3} +O(L^{4}) .\eqno (A.3)
$$
Thus, if the curve $\gamma$ is convex scattering, then the condition
$(\tan r(l))''\leq 0$ holds everywhere on $\gamma$.
Recall that $\tan r = \kappa^{-1}$.  If the strict inequality
$(\kappa^{-1}(l_{0}))''<0$ holds, then, by eq. (A.3), there is $L^{cr}$
such that the
arc $\gamma (l): l\in [l_{0}, l_{0}+L^{cr}]$ is convex scattering.
Thus, any sufficiently short curve satisfying the condition
$(\kappa^{-1})''<0$ is convex scattering.

By the choice of the coordinate system we have $|x(l)|\leq \max (r)$ for the corresponding quantities on $\gamma(l), \,\, l_0\leq l\leq l_1$.  Then,  we can obtain for $F(\theta(l),r(l),x(l)), \,\, l_0\leq l\leq l_1$ the estimate

$$F< (\tan r)_{max}( 1+3(\tan r)_{max}^2+ 5|(\tan r)'|_{max}) ,\eqno (A.4)$$
where $(\tan r)_{max}, |(\tan r)'|_{max}$ are the maxima  of the respective
quantities on $\gamma$ between the points $\gamma(l_0)$ and $\gamma(l_1)$.  Eq. (A.2)  implies that if  the inequality

$$ -(\tan r)'' > (\tan r)_{max}( 1+3(\tan r)_{max}^2+ 5|(\tan
r)'|_{max}),\eqno (A.5)$$
holds everywhere, then $\gamma $ is convex scattering.


\subsection* {B:  The hyperbolic plane}

Let  $\alpha$ and $\beta$ be a pair of geodesics in $\H^2$, intersecting
orthogonally. Just like in part A, we associate with this a coordinate system
$(x,y)$  on the hyperbolic plane.
For a convex curve, $\gamma$, and two points,
$\gamma (l_{0})$ and $\gamma (l_{1})$ of  $\gamma$,
we choose the geodesics $\alpha$ and $\beta$ like in part A, see fig. 14b.
Then the curvature $\kappa=\coth r$ of $\gamma$ satisfies
$$
{dx\over dl}=\cos\theta,  \qquad  {dy\over dl}={\sin\theta \over \cosh x}
,\eqno (B.1a)
$$
$$  {d\theta\over dl}=-\sin\theta \tanh x -\coth r ,\eqno (B.1b)$$
where $\theta (l)$ is the angle between the geodesic through the point $A$,
orthogonal to $\beta$, and $\gamma$. By straightforward calculations, we obtain
\begin{eqnarray*}\Delta&=&s-d_{1}-d_{2}=\int d(\arctan{\rm\! h}(\tanh
r\sin\theta ))+dx\\
         &=& \int dy{ \Big( (\tanh r )'- \cos\theta\tanh r(\tanh
x+\sin\theta\tanh r )\Big)
\cosh x \over 1-\tanh^{2} r \sin^{2}\theta}.\end{eqnarray*}
Set
\begin{eqnarray*}
 \lefteqn{F(\theta,r,x)} \\
 &=&-\tanh r\sin^{2}\theta (1+\tanh^{2} x)-\sin^{3}\theta\tanh
x\tanh^{2}r-\sin\theta\tanh x    \\
 &-&   (\tanh r )'\tanh r\sin 2\theta  
 +{(\tanh r )' - \cos\theta\tanh r(\tanh x+\sin\theta\tanh r )  \over
1-\tanh^{2} r \sin^{2}\theta}  \\
 &\times & \Big( (\tanh^{2} r)'\sin^{2}\theta +\tanh^{2} r \sin\theta\sin
2\theta\tanh x+\tanh r\sin 2\theta \Big).
\end{eqnarray*}
Then, since $y(l_{0})=y(l_{1})=0$, we have
$$\Delta=-\int dl\, \Big( (\tanh r )''+F(\theta(l),r(l),x(l)) \Big) {y\cosh
x \over 1-\tanh^{2} r\sin^{2}\theta}, \eqno (B.2) $$

Let $L=l_{1}-l_{0}$. By eq. (B.2), we obtain
$$\Delta (L)=-{(\tanh r)'' \over 12\tanh r}L^{3} + O(L^{4}),\eqno (B.3)$$
This leads to the necessary condition for convex scattering curve on the
hyperbolic plane: $(\kappa^{-1})''\leq 0$. Just like in part A,
eq. (B.3) implies that any sufficiently short arc satisfying
$(\kappa^{-1})''<0$ is convex scattering.

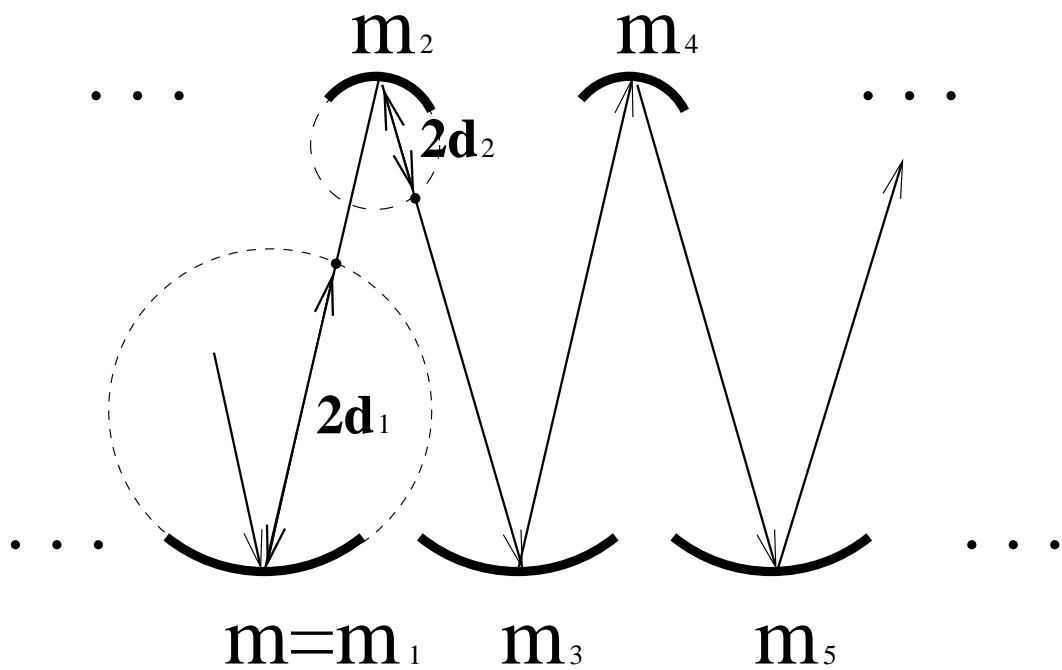
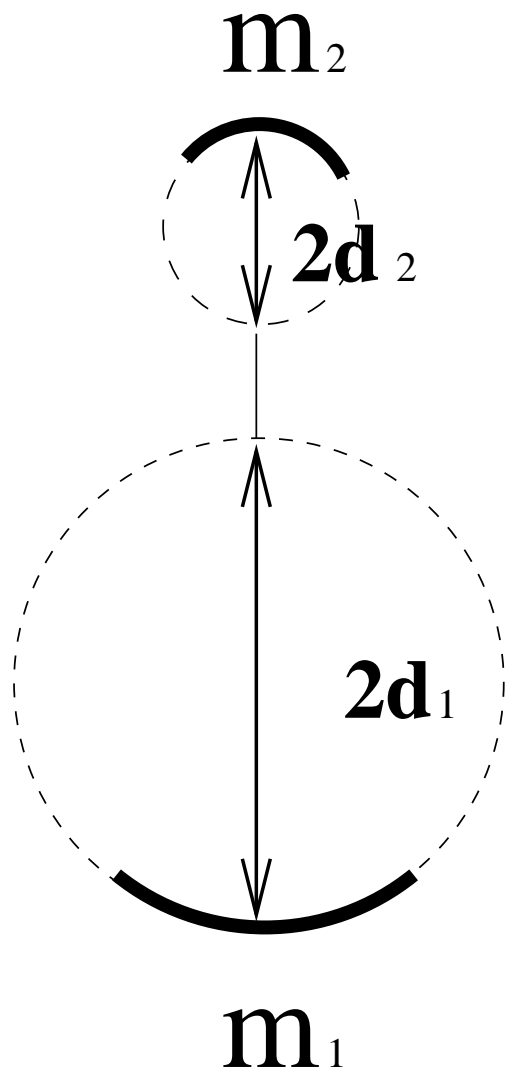
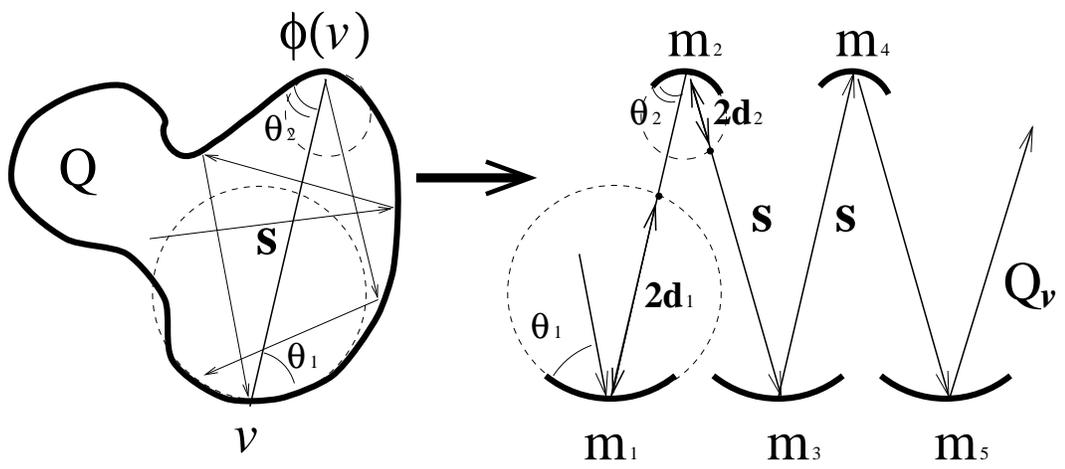
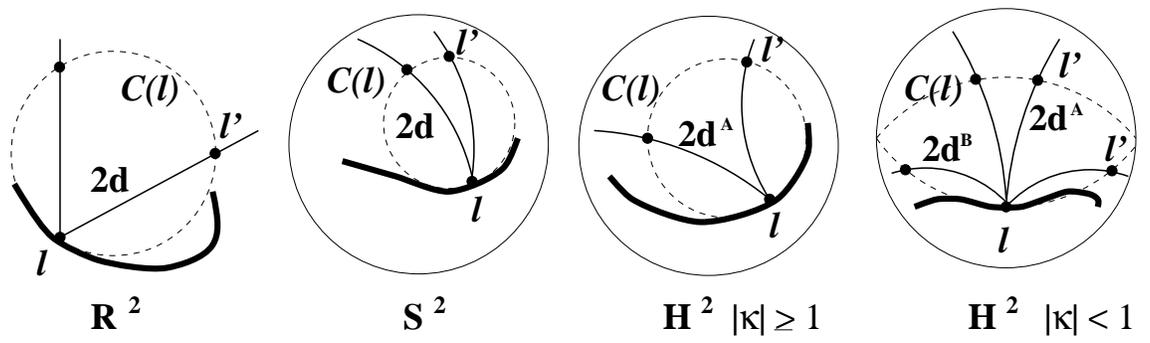
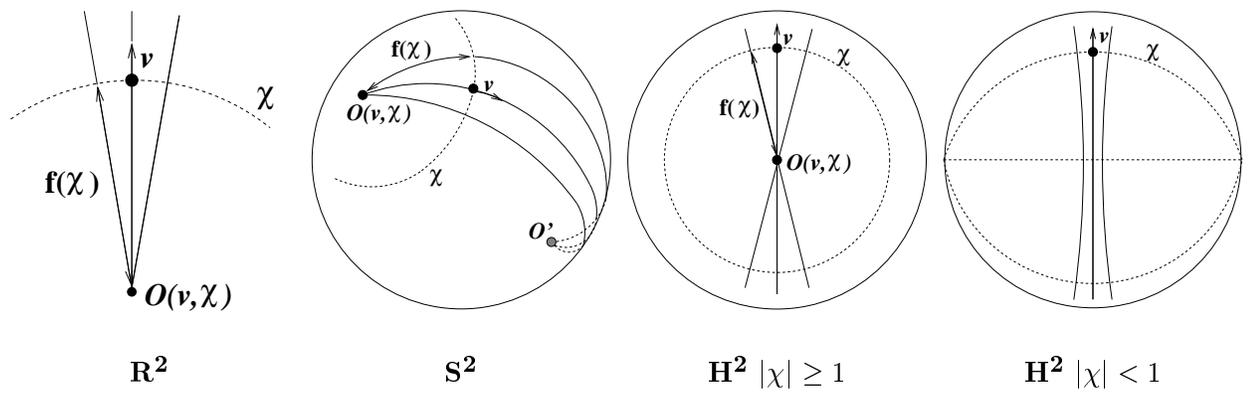
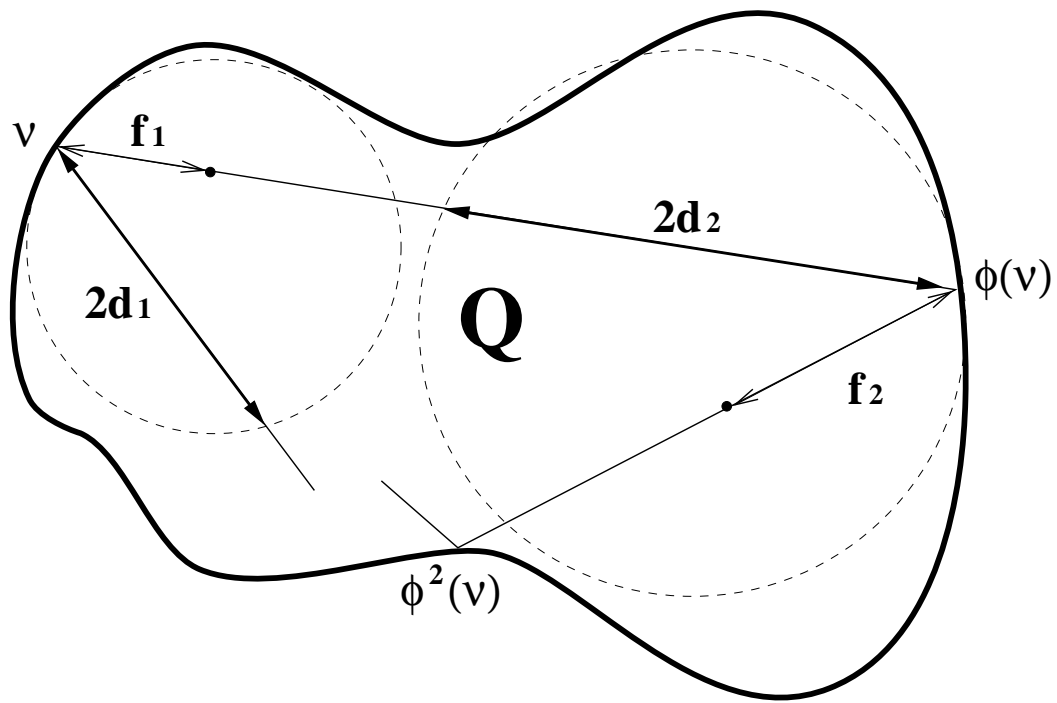
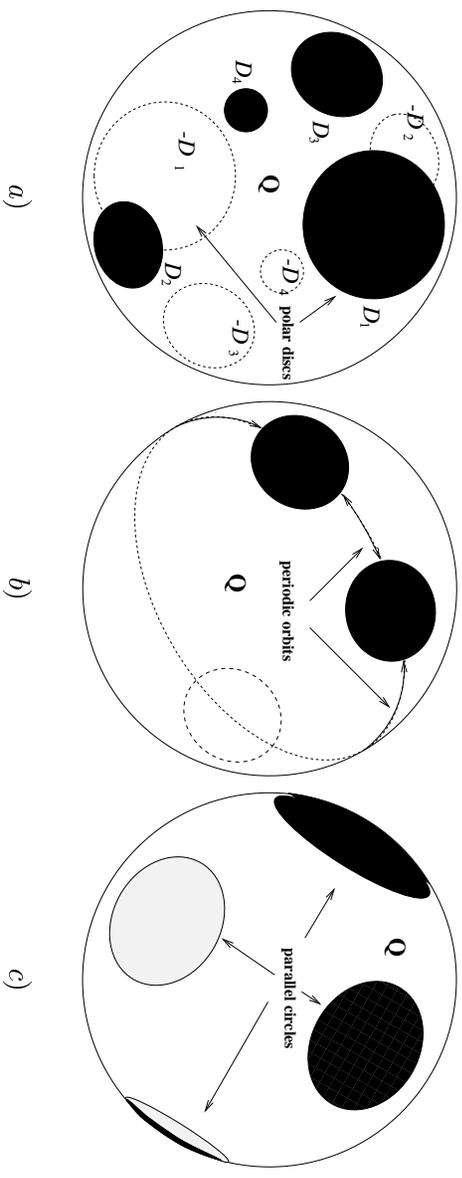
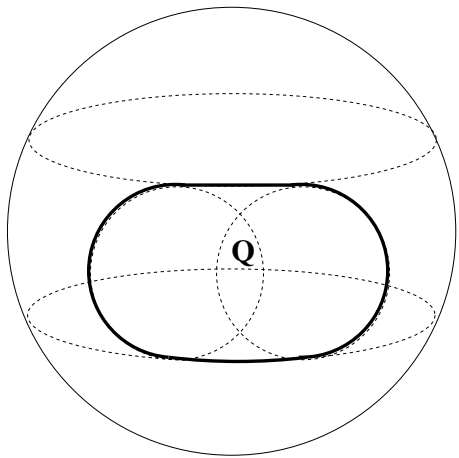
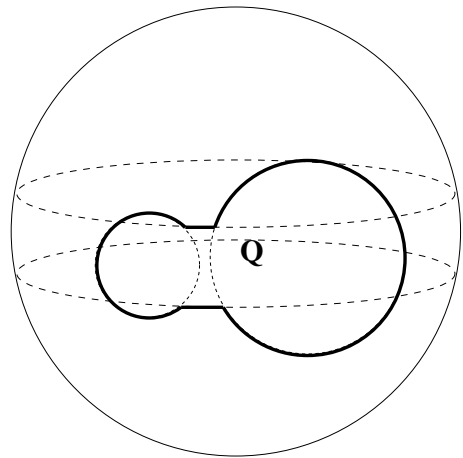
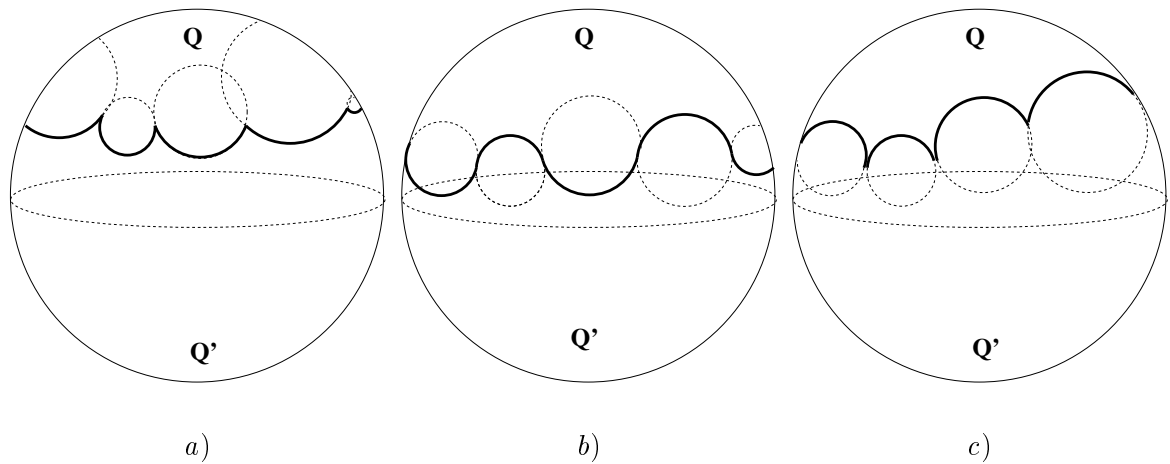
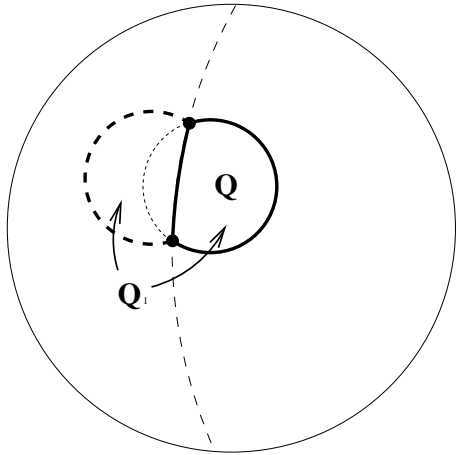
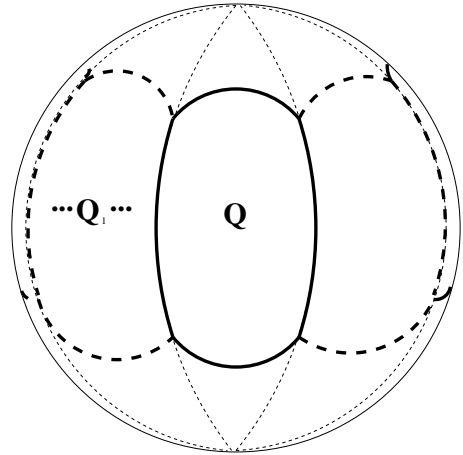
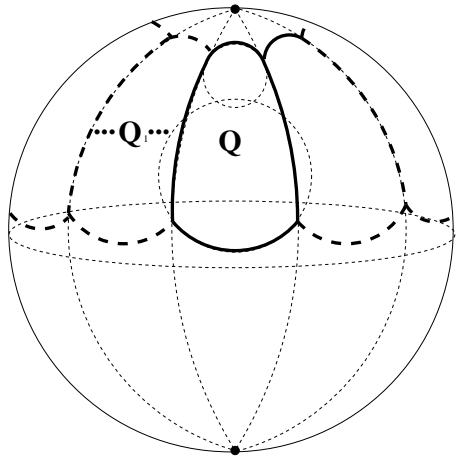
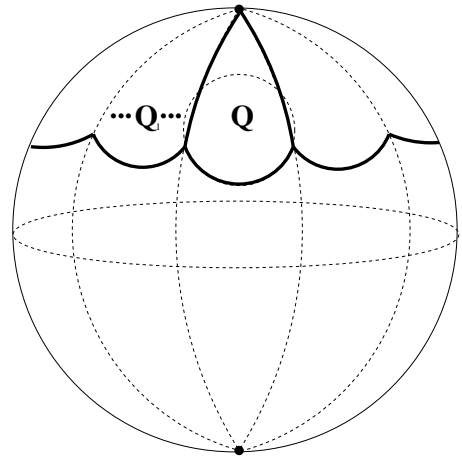
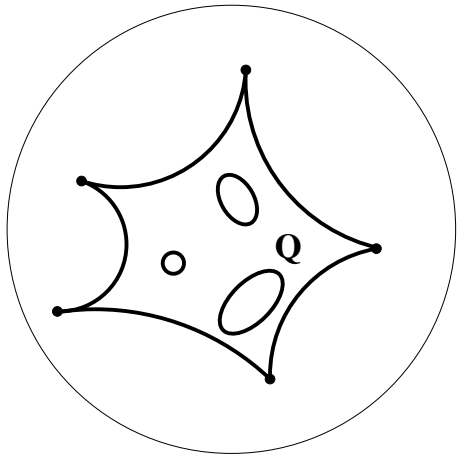
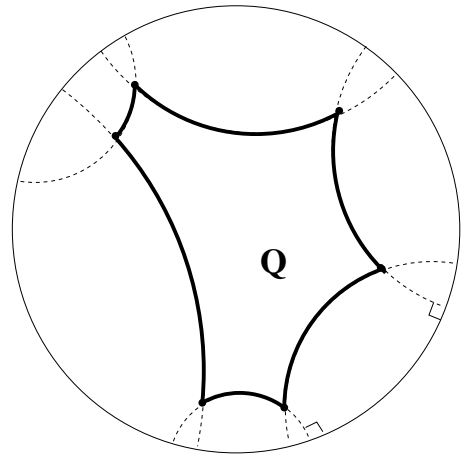
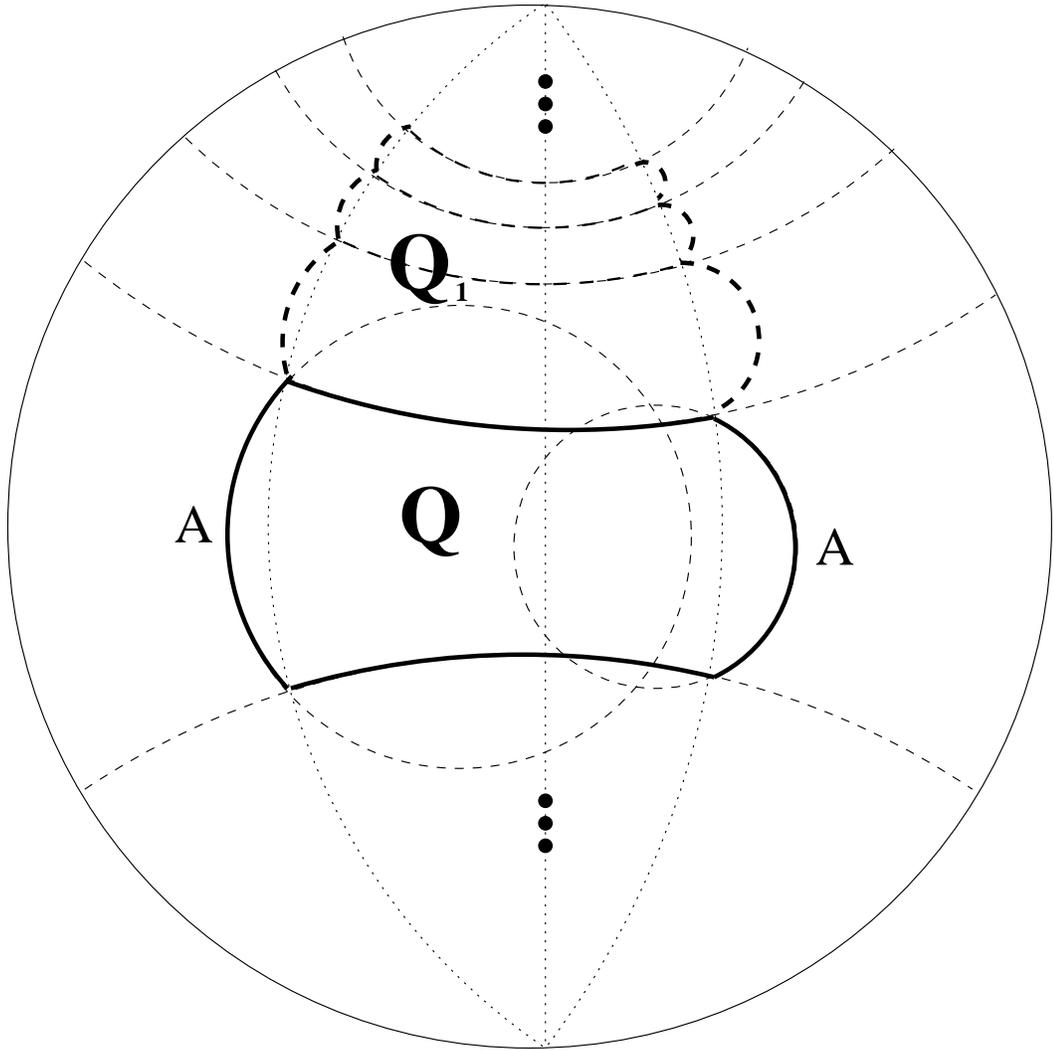
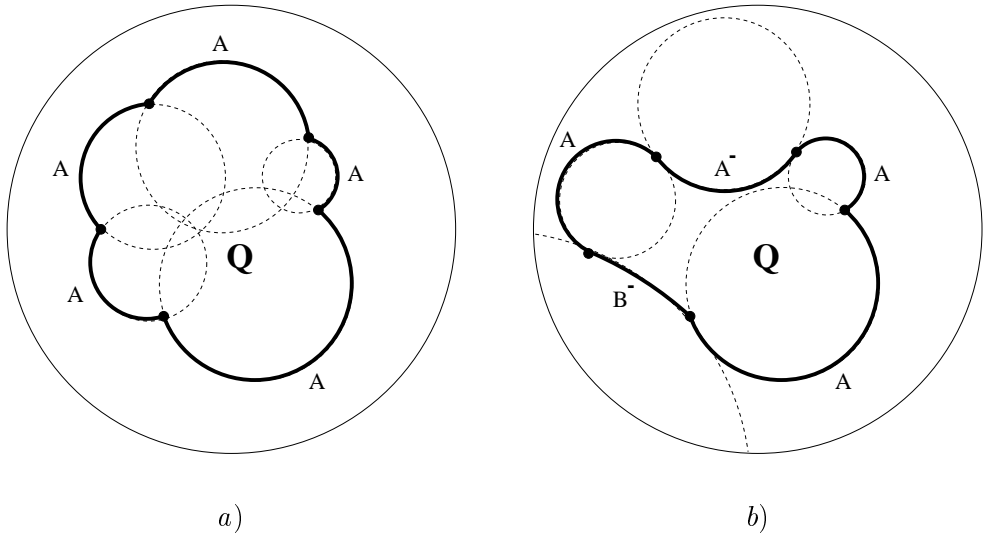
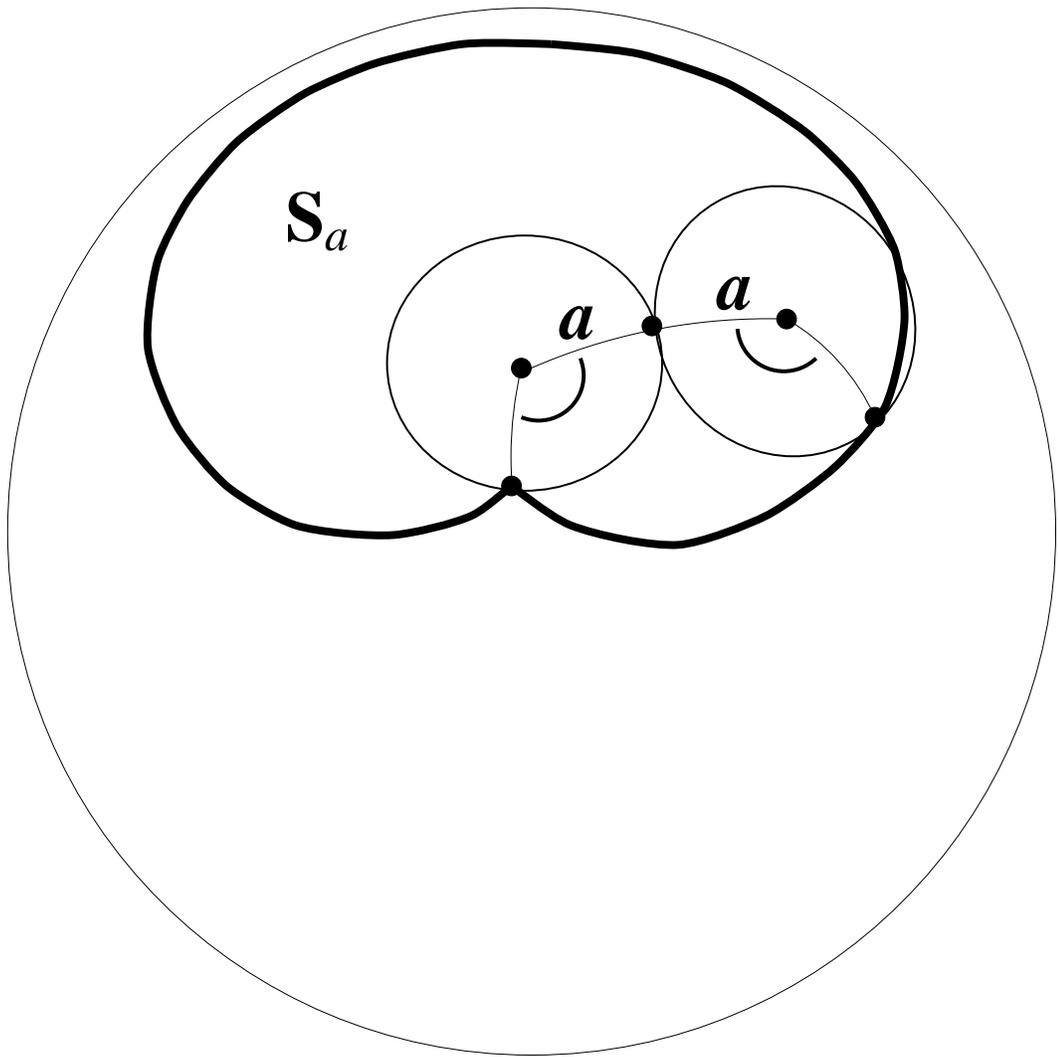
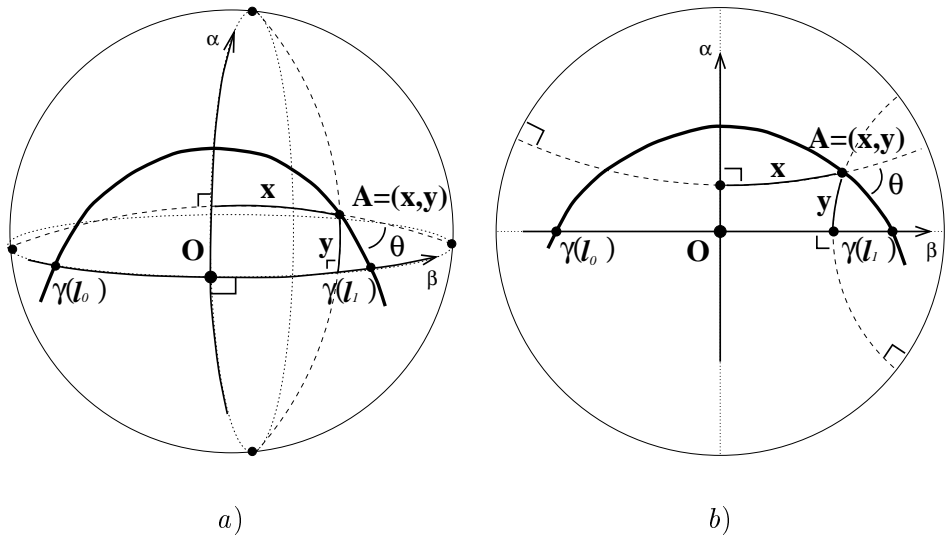
\end{document}